\documentstyle[11pt]{article}
\begin{document}
\begin{titlepage}
\begin{flushright}
IMPERIAL/TP/93-94/22 \\
astro-ph/9403006\\
(Submitted to {\bf Physical Review D})\\
\end{flushright}
\begin{center}
\LARGE
{\bf Numerical Methods In Cosmological Global Texture Simulations}\\
\vspace{1cm}
\normalsize
\large{Julian Borrill}\\
\normalsize
\vspace{.5cm}
{\em Blackett Laboratory,\\
Imperial College of Science and Technology,\\
Prince Consort Road, London SW7 2BZ,~~~U.~K.}\\
\vspace{.5cm}
\end{center}
\baselineskip=24pt
\begin{abstract}
\noindent
Numerical simulations of the evolution of a global topological defect
field have two characteristic length scales --- one macrophysical, of
order the field correlation length, and the other microphysical, of
order the field width. The situation currently of most interest to
particle cosmologists involves the behaviour of a GUT-scale defect
field at the epoch of decoupling, where the ratio of these scales is
typically of order $10^{50}$. Such a ratio is unrealisable in
numerical work, and we consider the approximations which may be
employed to deal with this. Focusing on the case of global texture we
outline the implementation of the associated algorithms, and in
particular note the subtleties involved in handling texture unwinding
events. Comparing the results in each approach then establishes that,
subject to certain constraints on the minimum grid resolution, the
methods described are both robust and consistent with one another.
\end{abstract}
\begin{center}
{\small PACS numbers: \hspace{0.5cm} 98.80.Cq, 98.70.Vc}\\
\end{center}
\end{titlepage}


\section{Introduction}

Topological defects as a class of phenomena naturally emerge in the
very early universe from the symmetry breaking phase transitions
associated with Grand Unifying extensions to the standard $SU(3)_{C}
\otimes SU(2)_{L} \otimes U(1)_{Y}$ model of particle physics
\cite{K}. A minimal requirement for any theory incorporating such
defects is that the energy density in the defect field does not
dominate the universe today. In certain cases the defect field is
believed to evolve to a scaling regime, wherein its energy density
remains a fixed small fraction of the total. Such defects can then be
invoked as a possible source of the primordial density perturbations
believed to have seeded cosmological structure formation and observed
as fluctuations in the cosmic microwave background (CMB) by the COBE
satellite \cite{COBE}. The essential observational constraints on any
model of structure formation based on topological defects --- most
notably invoking either gauged cosmic string \cite{STRING} or global
texture \cite{TEXTURE} --- are therefore of the large scale structure
of the local universe essentially as it is now, at a time $t \sim
10^{18}$ s after the big bang, and of the CMB radiation from the epoch
of matter/radiation decoupling at $t \sim 10^{13}$ s. By contrast the
assumed defect-forming Grand Unified Theory symmetry breaking
typically occurs at $t \sim 10^{-35}$ s. Numerical simulations of the
evolution and effects of the defect field have therefore come to play
an essential role in bridging this gap of some 50 orders of magnitude
in time.

A crucial factor determining the type of numerical simulation used is
whether the symmetry broken at the phase transition is gauged or
global. For a broken gauged symmetry any vacuum field configuration
which is not topologically stablised can simply be trivialised by a
suitable gauge transformation, and all that need be considered is the
behaviour of the defects themselves which can be treated as discrete
objects. For a broken global symmetry however topologically unstable
vacuum field configurations are only trivialised by correlation as the
field evolves. Since the correlation length of a global defect field
is typically taken to be of order the horizon scale its evolution must
be followed in detail up to this scale. Cosmological global defect
field simulations thus contain two fundamental length scales --- one
macrophysical, of order the horizon size, and the other microphysical,
of order the inverse mass of the field --- which we can quantify via
their ratio as
\begin{equation}
{\cal R} \equiv \frac{L_{macro}}{L_{micro}} \sim
\frac{H^{-1}}{M_{\phi}^{-1}}
\end{equation}
For a GUT-scale defect at the epoch of decoupling ${\cal R} \sim
10^{50}$ and our characteristic length scales again differ by 50
orders of magnitude. Since it is not yet possible to simulate such a
range of scales simultaneously we must adopt one of two possible
approximation schemes. One approach is to reduce the field's mass
scale, and hence increase its width, to a level which is numerically
resolvable \cite{PRS,STPR,BCL1,BCL2}. Since the field's width
characterises the size of the defect core this may be dubbed the
expanded core (XCORE) approximation. The alternative is to take the
field's mass scale to infinity and hence its width to zero, removing
the microphysical scale from the problem \cite{BR,PST,CPT,DS/DHZ}.
Provided this is done self-consistently the field's equations of
motion then reduce to those of the non-linear sigma model (NLSM).

In this paper we consider the implementation of numerical simulations
of the evolution of global texture in each of these approaches. In
each case the approximated field mass is very different from the
physical mass, and it is essential to determine the conditions under
which the resulting simulated evolution still reflects the true
behaviour of the field. In the XCORE case some progress has already
been made \cite{B93} by finding the minimal grid resolution necessary
for some of the properties characterising the field's evolution to
asymptote. Here, however, we are able to go further; since the two
approaches involve making diametrically opposed approximations
consistency between them would be strong evidence that each is close
to the true field evolution. Our primary focus therefore is on
comparing results from each simulation approach and testing their
mutual consistency.

As well as being one of the more promising candidate defect models,
global texture also displays some unique features in its evolution. In
particular although the texture field is globally topologically
trivial, being everywhere on the vacuum manifold with no regions
trapped in the false-vacuum phase, it nonetheless admits
configurations which are in some sense {\it locally} topologically
non-trivial. In more than two spatial dimensions such configurations
are unstable, with their field gradient energy concentrating (by
analogy with Derrick's theorem \cite{D}) until the local energy
density is sufficient to pull the field off the manifold and through
the false-vacuum, `unwinding' this locally non-trivial topology. Since
such unwinding events, where the field leaves the vacuum manifold, are
where the different approximation schemes are most likely to diverge,
they will provide the most stringent test of consistency.

Having established our notation we commence by detailing the continuum
form and discretised implementation of the equations of motion in each
approximation scheme. We include in this some discussion of the
identification of unwinding events and an analysis of the minimal grid
resolution necessary for the approximations to model such events
accurately. Working within these constraints we are then able, at
least in flat space, to check the simulations against one another. For
a number of initial field configurations, both with and without
unwinding events, we consider the field's point-by-point evolution,
its gross features, and the CMB anisotropies it generates in each
approximation.

\section{Notation And Conventions}

Except where specified otherwise we consider the evolution of a
4-component real scalar global field $\Phi$ in $3 + 1$ dimensional
flat space with metric signature $(-,+,+,+)$. This field evolves in a
potential of the form
\begin{equation}
V(\Phi) = V_{\rm o} (|\Phi|^{2} - \Phi_{\rm o}^{2})^{2}
\end{equation}
which is defined such that above some critical temperature $\Phi_{\rm
o}^{2} < 0$ and the potential has an $O(4)$ symmetry about its minimum
at $\Phi = 0$, whilst below it $\Phi_{\rm o}^{2} > 0$ and the symmetry
is broken to $O(3)$ about the new minimum which lies somewhere on the
entire $S_{3} = O(4)/O(3)$ manifold of degenerate vacuum states
defined by $|\Phi| = \Phi_{\rm o}$. This is the simplest symmetry
breaking to generate global texture.

In discussing texture configurations it is often useful to
parameterise the field at some point in space given by the spherical
polar co-ordinates $(r, \theta, \psi)$ at some time $t$ as
\begin{equation}
\label{ePARAM}
\Phi(r, \theta, \psi, t) = |\Phi| (\cos {\cal X}, \sin {\cal X}
\cos \Theta, \sin {\cal X} \sin \Theta \cos \Psi, \sin {\cal X} \sin
\Theta \sin \Psi)
\end{equation}
where the group space angles $0 \leq {\cal X} < \pi$, $0 \leq \Theta <
\pi$ and $0 \leq \Psi < 2 \pi$ locate the projection of $\Phi$ on
the vacuum manifold. In general ${\cal X}$, $\Theta$ and $\Psi$ are
functions of position and time, although spherically symmetric field
configurations can be realised in a particularly simple way by
identifying two of the group space angles with the two physical space
angles, $\Theta \equiv \theta$ and $\Psi \equiv \psi$.

For convenience we further normalise the field, $\Phi \rightarrow
\Phi/\Phi_{\rm o}$. All simulations are now performed on lattices with
grid spacing $\delta x$ and using time steps $\delta t$ and with the
speed of light set to $c=1$ so that grid distances and times are
numerically interchangeable. The discretised field is labelled
$\Phi_{i,n}$ with i-indices spatial and n-indices temporal.

\section{Implementation}

\subsection{Equations Of Motion}

The full texture Lagrangian
\begin{equation}
{\cal L} = - \frac{1}{2} \partial^{\mu} \Phi \partial_{\mu} \Phi -
V_{\rm o} (|\Phi|^{2} - 1)^{2}
\end{equation}
gives the equations of motion of the field as
\begin{equation}
\Box \Phi = - \frac{\partial V}{\partial \Phi}
\end{equation}
In flat space these reduce to
\begin{equation}
\label{eXEQ}
\ddot{\Phi} - \nabla^{2} \Phi = -4 V_{\rm o} (|\Phi|^{2} - 1) \Phi
\end{equation}
Using a standard staggered leapfrog method \cite{PFTV} these can be
discretised to second order to yield
\begin{eqnarray}
\dot{\Phi}_{i,n+1/2} & = & \dot{\Phi}_{i,n-1/2} + \left(\nabla^2 \Phi_{i,n}
       - 4 V_{\rm o} (\Phi^{2}_{i,n} - 1) \Phi_{i,n} \right) \delta t
\nonumber \\
\Phi_{i,n+1} & = & \Phi_{i,n} + \dot{\Phi}_{i,n+1/2} \delta t
\end{eqnarray}
These are the equations used in the XCORE approximation.

Alternatively, imposing the NLSM approximation and restricting the
field to remain on the vacuum manifold at all times, the constrained
Lagrangian can be written
\begin{equation}
{\cal L} = - \frac{1}{2} \partial^{\mu} \Phi \partial_{\mu} \Phi -
\frac{\sigma}{2} (|\Phi|^{2} - 1)
\end{equation}
where $\sigma$ is a Lagrange multiplier. The equations of motion now
become
\begin{equation}
\Box \Phi = \sigma \Phi
\end{equation}
giving
\begin{equation}
\sigma = (\Box \Phi) \cdot \Phi
\end{equation}
Using the constraint to solve this equation for $\sigma$ the flat
space equations become
\begin{equation}
\ddot{\Phi} - \nabla^{2} \Phi = - (\dot{\Phi}^{2} - \nabla \Phi^{2})
\Phi
\end{equation}
Again discretising to second order these give
\begin{equation}
\Phi_{i,n+1} = 2 \Phi_{i,n} - \Phi_{i,n-1} + \nabla^2 \Phi_{i,n} \delta
t^2 - (\dot{\Phi}_{i,n}^2 - \nabla \Phi_{i,n}^2)\delta t^2
\Phi_{i,n}
\end{equation}
Following Pen et al \cite{PST} this can be cast in the form
\begin{equation}
\label{eNEQ}
\Phi_{i,n+1} = \lambda \Phi_{i,n} + \delta \Phi_{i,n}
\end{equation}
where
\begin{eqnarray}
\delta \Phi_{i,n} & = & \Phi_{i,n} - \Phi_{i,n-1} + \nabla^2
\Phi_{i,n} \delta t^2 \nonumber\\
\lambda & = & 1 - (\dot{\Phi}_{i,n}^2 - \nabla \Phi_{i,n}^2)\delta t^2
\end{eqnarray}
Using the constraint $|\Phi_{i,n+1}| = 1$ to solve for $\lambda$ now
gives
\begin{equation}
\label{eLAM}
\lambda = - \delta \Phi_{i,n}.\Phi_{i,n} \pm \sqrt{1 - \delta
\Phi_{i,n}^{2} + (\Phi_{i,n}.\delta \Phi_{i,n})^2}
\end{equation}
and we are able to circumvent the numerically troublesome squared
derivative terms. Note that to first order in $\delta t$ equation
(\ref{eNEQ}) is
\begin{equation}
\Phi_{i,n+1} = \pm \Phi_{i,n} + {\rm O}(\delta t)
\end{equation}
with the choice of sign coming from the choice of root in equation
(\ref{eLAM}); taking the negative root, so that $\Phi_{i,n+1} \sim -
\Phi_{i,n}$, therefore explicitly introduces an unwinding event at the
given gridpoint.

\subsection{Unwinding Events}

Evolving the texture field on a discrete lattice there are two types
of unwinding event. If the initial field conditions are sufficiently
symmetric about some gridpoint that the group space angles there are
time-independent then an unwinding can occur at that gridpoint.
However for more random initial field conditions the unwinding site
will not lie exactly at a gridpoint. To investigate the properties of
individual texture unwindings, as well as to test the accuracy of our
simulations generally, we would like to be able to identify both types
of event unambiguously.

The full texture field equations of motion include unwinding events
dynamically. Reducing the mass scale of the field allows it to be
pulled off the vacuum manifold more easily, so that making the XCORE
approximation increases the volume of space leaving the vacuum and
forming the defect core. The constraint on the field mass is then that
it should be sufficiently small that any unwinding event will pull the
field below some threshold value --- typically taken as $|\Phi_{i,n}|
< 0.5$ --- throughout a volume containing at least one gridpoint.
Conversely the mass must also be large enough that the field is only
pulled below this threshold at unwinding events. However, since we are
trying to maximise the range of length scales, and hence require as
large a mass as possible, only the upper limit serves as a serious
restriction. A further constraint on the field's mass comes from the
upper limit on field gradients imposed by the finite grid spacing; if
the width is too small then even unwindings occuring at gridpoints can
be inhibited by the resulting upper limit on the field's gradient
energy density. A detailed investigation of these constraints \cite{B93}
leads to a practical lower limit on the field width
\begin{equation}
W \geq 0.25 \, \delta x
\end{equation}
Subject to this, and the grid resolution conditions discussed below,
the threshold prescription serves to identify unwindings in the XCORE
approximation.

Since the NLSM approximation constrains the field to remain on the
vacuum manifold at all times it is unable to admit unwinding events in
the continuum. However, as demonstrated above, the discretised
equations of motion include a means by which an unwinding can be
explicitly re-introduced at a given gridpoint. All that remains is to
determine the appropriate criteria under which such an unwinding
should be incorporated in the field's evolution. For diagramatic
simplicity in the remainder of this section we consider the case of a
3-component NLSM texture field in $2 + 1$ dimensions where the vacuum
manifold is $S_{2}$.

The most obvious approach is to introduce an unwinding at any
gridpoint in the immediate vicinity of which the field covers more
than half of the vacuum manifold --- ie. if we consider the bisection
of the manifold by the closed boundary on it given by joining the
nearest neighbours to any gridpoint along geodesics then we introduce
an unwinding whenever the field at the gridpoint lies in the larger of
the two sub-sections. To date the only attempt to implement this
criterion \cite{PST,CPT} involves calculating the mean value of the
(here 4) nearest neighbours to a gridpoint, $\bar{\Phi}_{i,n}$, and
introducing an unwinding if and only if $\Phi_{i,n} \cdot
\bar{\Phi}_{i,n} < 0$. This is equivalent to replacing the actual
boundary on the manifold with an `averaged' one defined by the
intersection of the manifold with the plane perpendicular to
$\bar{\Phi}_{i,n}$ (figure 1). Note that without loss of generality we
can always take $\bar{\Phi}_{i,n}$ to point `south', whereupon
$\Phi_{i,n} \cdot \bar{\Phi}_{i,n} < 0$ if and only if $\Phi_{i,n}$
and $\bar{\Phi}_{i,n}$ lie in opposite hemispheres.

In the symmetric case it is straightforward to show that
\begin{equation}
\bar{\Phi}_{i,n} \propto \Phi_{i,n}
\end{equation}
and hence that $\Phi_{i,n}$ and $\bar{\Phi}_{i,n}$ lie in opposite
hemispheres if and only if more than half of the vacuum manifold is
covered locally (figure 2).

For non-symmetric configurations this approach breaks down in a number
of ways:

\begin{itemize}

\item since in general $\bar{\Phi}_{i,n} \not\propto \Phi_{i,n}$ it is
simple to construct configurations where $\Phi_{i,n}$ and
$\bar{\Phi}_{i,n}$ lie in the same hemisphere but which should
nonetheless be unwound (figure 3). As is clear from this example the
correct condition for identifying an unwinding is instead
$(\Phi_{i,n} - \bar{\Phi}_{i,n}) \cdot \bar{\Phi}_{i,n} < 0$.

\item if the field at the gridpoint in question lies between the true
boundary and the `average' boundary then even the correct condition
either mistakenly identifies or fails to identify an unwinding
configuration (figure 4).

\item since in general the unwinding site is not at a gridpoint even
if the algorithm correctly identifies a configuration that should
unwind what we are doing in effect is unwinding the gridpoint closest
to the actual unwinding site. However in the NLSM approximation {\it
only} at the unwinding site itself should the field ever leave the
vacuum manifold. Moreover, even if the NLSM condition is relaxed and a
small region of space close to the unwinding site (and including the
gridpoint) is allowed to leave the manifold it is only at the
unwinding site itself that the field is approximately inverted,
$\Phi_{i,n+1} \sim -\Phi_{i,n}$. This approach therefore incorrectly
evolves the field at the gridpoint, and can thereby generate a
repeated unwinding (figure 5).

\end{itemize}

In practice unless the unwinding site is fixed at a gridpoint (when
the required symmetry avoids the above difficulties) it is unnecessary
to explicitly unwind the field at all, since the geodesic assumption
(that between neighbouring points the field follows a geodesic on the
manifold) implicitly unwinds any configuration locally covering more
than half the manifold anyway. However the problem of unambiguously
identifying unwindings in the NLSM approximation in order to study
their properties remains unsolved.

\subsection{Grid Resolution}

Given the above microphysical length scales, the resolution afforded
by a particular simulation lattice can be quantified by the number of
gridpoints $N_{l}$ corresponding to the initial macrophysical length
scale. For comparison between different approaches we need a length
scale which is well-defined and which can be meaningfully implemented
in all simulations --- for example in flat space the horizon is not
well-defined, whilst symmetric initial field configurations have an
infinite correlation length.

A necessary condition for any field configuration to unwind that to do
so must reduce the gradient energy in the configuration, and hence
that it should cover more than half of the vacuum manifold before
unwinding, and by corollary less than half afterwards. Therefore, as
in previous work \cite{BCL1,BCL2}, we adopt as our macrophysical
length scale the radius of a sphere within which no more than half of
the manifold can possibly be covered initially. For singly-wound
spherically symmetric configurations \cite{BCL1} the manifold covering
about the origin is given (in the ansatz of equation (\ref{ePARAM}))
by
\begin{equation}
Q(r) = \frac{1}{\pi} \left( {\cal X}(r) -
\frac{1}{2} \sin 2 {\cal X}(r) \right)
\end{equation}
and our length scale is the radius at which $Q = 0.5$ initially. For
interpolated random configurations \cite{BCL2} it is half the distance
between gridpoints at which the initial field values are randomly
assigned.

Figure 6 illustrates the variation in the topological charge causally
associated with the unwinding of a spherically symmetric configuration
in flat space $Q(t_{uw}-t_{\rm o})$ (where $t_{uw}$ and $t_{\rm o}$
are the unwinding and initial times respectively), with the grid
resolution $N_{l}$ in each approximation.  To get within 10\% of the
asymptotic result in both simulations simultaneously a lower limit on
the number of gridpoints per half-manifold radius in flat space is
$N_{l} \geq 5$, and to within 5\%, $N_{l} \geq 10$. However it should
be stressed that these results {\it only} apply to spherically
symmetric configurations in flat space. Experiments using the XCORE
approximation in expanding backgrounds and with random configurations
demonstrate that in such circumstances the minimum resolution is
increased \cite{B93}, and wherever possible the simulations presented
here satisfy the tighter constraint from that work, $N_{l} \geq 16$.

\section{Comparison}

The evolution of an initial field configuration under each of the
approximations can be compared in a number of ways; here we consider
three levels of consistency. Firstly we can calculate point by point
statistics measuring the correlation of the field over the lattice at
every time step. Alternatively we can consider the consistency of any
gross features of the simulations, such as the times and locations of
any identifiable unwinding events. Finally we can compare observable
consequences of the field evolutions, such as their induced CMB
anisotropies. Clearly these levels of consitency are hierarchical; if
the simulations are consistent point by point then their gross
features and physical consequences will be identical. Similarly even
if their point by point evolution varies, provided the gross features
are the same then we might expect the induced CMB anisotropies to be
so too.

\subsection{Pointwise Statistics}

Given that the vacuum manifold is the 3-sphere $|\Phi| = 1$ a simple
measure of the correlation of the field at any point and time is given
(in an obvious notation) by
\begin{equation}
\xi_{i,n} = \Phi^{\mbox{\tiny X}}_{i,n} \cdot \Phi^{\mbox{\tiny N}}_{i,n}
\end{equation}
However we know at the outset that differences will emerge as the
simulations progress simply because the field is able to leave the
vacuum manifold in the XCORE but not in the NLSM approximation. As a
further comparison therefore we can factor out this effect by
comparing the normalised fields, projecting the XCORE field back onto
the vacuum manifold everywhere
\begin{equation}
\hat{\xi}_{i,n} = \frac{\Phi^{\mbox{\tiny X}}_{i,n}}
{|\Phi^{\mbox{\tiny X}}_{i,n}|} \cdot \Phi^{\mbox{\tiny N}}_{i,n}
\end{equation}
In this way we can determine the relative importance in any absence of
correlation of differences in the field positions' magnitudes and of
differences in their orientations.

Symmetric configurations are artificially self-correlated on all
scales at all times, possibly enhancing their cross-correlation too,
so we focus on the evolution of random configurations alone. Using a
catalogue of random initial configurations admitting unwinding events
\cite{BCL2} we take a sample of 22 configurations, equally divided
between ones with and ones without an unwinding. These are then
simultaneously evolved on a pair of $48^3$ lattices for 24 grid time
units. Running the simulations simultaneously restricts us to half the
usual grid size (and hence resolution) here, so the results should be
taken as upper limits which would be reduced were it possible to work
at full resolution. We calculate as our correlation statistic the mean
values of each measure over the simulation lattice at each timestep
\begin{equation}
\xi_{n} = \frac{1}{N} \sum_{i=1}^{N} \xi_{i,n}
\;\;\;\;\;\;\;\;\;\;\;\;\;\;\;\;\;\;
\hat{\xi}_{n} = \frac{1}{N} \sum_{i=1}^{N} \hat{\xi}_{i,n}
\end{equation}
Exact correlation gives a statistic of 1, anti-correlation -1, and no
correlation 0. In all simulations and at every timestep we find
$\xi_{n} = \hat{\xi}_{n} = 1.0$.

To test of the degree of this consistency we also note the number and
distribution of gridpoints at which either of the correlation measures
falls below 0.9. In simulations of both unwinding and non-unwinding
configurations gridpoints whose fields are less than 90\% correlated
are found localised in time and space about events during which the
field is pulled significantly off the manifold in the XCORE
simulations. This is true both for the known unwinding events, and for
the non-unwinding events (at which insufficient gradient energy
collapses to unwind the field) which may be identified {\it post hoc}
by the characteristic CMB anisotropy pattern they induce \cite{BCLSV}.
However, these imperfectly correlated regions are very small; even the
largest region involves less than 0.1\%, and a typical region less
than 0.01\%, of the simulation gridpoints.

Figure 7 shows the variation in the number of gridpoints whose fields
are less than 90\% correlated under each measure as the simulation
progresses for a typical configuration, in this case including an
unwinding at grid time 13.9. We see that for the full field the number
of imperfectly correlated points peaks just before and after the
unwinding event, but that only a small fraction of this discrepancy is
due to differences in orientation, and that these differences peak at
the unwinding itself. The sequence of post-unwinding peaks in the full
field correlation measure is to be expected given the suppression of
the damping of radial oscillations by the reduction in the field's
mass in the XCORE approximation. Note that, except for the first,
these peaks involve a single point; explicit examination of the
simulation confirms that this is indeed the site of the unwinding
event.

\subsection{Gross Field Features}

As shown above the only even marginally uncorrelated sections of the
simulations are associated with unwinding or non-unwinding events. We
now consider the extent to which the gross features of these events,
such as their time and place, reflect this and differ between the
simulations. The cross-correlation results already limit the possible
differences; in each case the event must occur within the imperfectly
correlated region of simulation spacetime, so their locations cannot
differ by more than the size of this region.

As a more precise test we can attempt to identify and thereby locate
each simulations' unwinding and non-unwinding events explicitly. The
unwinding events are identified in the XCORE simulation as occuring at
the gridpoint at which the magnitude of the field falls below some
threshold, $|\Phi| < 0.5$, and at the time half way between its first
falling below the threshold and subsequently exceeding it again. In
the NLSM simulations we use the criterion discussed above,
$(\Phi_{i,n} - \bar{\Phi}_{i,n}) \cdot \bar{\Phi}_{i,n} < 0$, whilst
acknowledging its weaknesses. Note that the criterion is used simply
to identify candidate unwinding events, and the equations of motion
are unchanged, with the positive root being taken in equation
(\ref{eLAM}). Where unwindings are successfully identified (in all but
one case) they are found to occur at the same gridpoint and within 0.5
grid time units of their XCORE counterparts.

Identification of non-unwinding events is less precise; in both XCORE
and NLSM simulations we look for the characteristic CMB signature ---
a sharply peaked cold spot followed by a similar hot spot
\cite{BCLSV}. By analogy with unwindings, the non-unwinding event is
then taken to occur at the gridpoint around which the peaks are
centred and at the time at which the cold to hot spot transition
occurs. In every case the event occurs at the same gridpoint and on
the same photon sheet. Since the sheets are separated by 1 grid
spacing this specifies the event's time again to within 0.5 grid time
units.

\subsection{CMB Anisotropies}

Finally we compare the CMB anisotropies induced by the texture field's
evolution in each simulation. The details of the method are described
elsewhere \cite{BCLSV}; in essence we send a sheet of photons across a
lattice over which the texture field is evolving and calculate the
`kick' given to each photon at each time step in the stiff source,
small angle approximation \cite{S}. The variation in the CMB
temperature across the sheet is then recovered by Fourier analysis.

Here we are interested in the difference in the temperature
anisotropies induced by a given initial field configuration evolved in
each approach; we calculate
\begin{equation}
\delta T = \frac{1}{N} \sum_{i=1}^{N}
\left((\Delta T/T)^{\mbox{\tiny X}}_{i} -
(\Delta T/T)^{\mbox{\tiny N}}_{i}\right)
\end{equation}
where the sum is over a given sheet of photons. For all configurations
and for every sheet $\delta T = 0$. However the standard deviation of
the temperature difference across the sheets clearly differs between
the unwinding and non-unwinding configurations. Taking $\epsilon = 8
\pi^{2} G \Phi_{\rm o}^{2}$, the magnitude of the anisotropy generated
by the one known analytic solution to the flat space NLSM equations
\cite{TS}, we find
\begin{equation}
\sigma \leq \left\{ \begin{array}{ll}
	0.05 \;\epsilon & \mbox{unwinding} \\
	0.01 \;\epsilon & \mbox{non-unwinding}
	\end{array} \right.
\end{equation}

Since the characteristic CMB anisotropy generated by an unwinding or
non-unwinding event involves a sharply peaked cold spot being
succeeded by a similar hot spot we can also consider differences in
the maximum and minimum anisotropies between simulation pairs. We
calculate
\begin{equation}
\delta T_{\rm max} = \frac{1}{N} \sum_{i=1}^{N}
\left((\Delta T/T)^{\mbox{\tiny X}}_{\rm max} -
(\Delta T/T)^{\mbox{\tiny N}}_{\rm max}\right)
\end{equation}
and
\begin{equation}
\delta T_{\rm min} = - \frac{1}{N} \sum_{i=1}^{N}
\left((\Delta T/T)^{\mbox{\tiny X}}_{\rm min} -
(\Delta T/T)^{\mbox{\tiny N}}_{\rm min}\right)
\end{equation}
where the sum is now taken over the set of all unwinding or all
non-unwinding simulations, and now find
\begin{eqnarray}
\delta T_{\rm max} & = & \left\{ \begin{array}{ll}
	(-0.02 \pm 0.04) \;\epsilon & \mbox{unwinding} \\
	(-0.03 \pm 0.03) \;\epsilon & \mbox{non-unwinding}
	\end{array} \right. \nonumber \\
\delta T_{\rm min} & = & \left\{ \begin{array}{ll}
	(-0.04 \pm 0.04) \;\epsilon & \mbox{unwinding} \\
	(-0.02 \pm 0.02) \;\epsilon & \mbox{non-unwinding}
	\end{array} \right.
\end{eqnarray}
where the error bars are $1 \sigma$. Thus it appears that on average
the NLSM gives marginally stronger peaks than the XCORE simulations,
although in all cases the difference is within $1 \sigma$. Indeed the
only consistent discriminant between the anisotropies generated in
each simulation is that the difference in individual peak heights
$(\Delta T/T)_{\rm max} - (\Delta T/T)_{\rm min})$ is in every case
slightly greater in the NLSM simulation than its XCORE counterpart,
although never by more than $0.1\;\epsilon$.

\section{Conclusions}

We have detailed the implementation of two approaches to circumventing
the problem of the range of scales inherent in numerical simulations
of global defect fields. It is to be expected that each approach will
be constrained by a minimum lattice resolution, and that such
constraints will be strongest whenever significant events occur on
microphysical length scales. In the case of global texture such events
are the unwindings and non-unwindings associated with the local
concentration of field gradient energy. We find that there is indeed a
minimum resolution --- of the order of 5 grid spacings per initial
half-manifold radius --- for such events to be modelled accurately for
spherically symmetric simulations in flat space.

For non-symmetric field configurations we have been unable to improve
on the previously established resolution limits \cite{B93} because of
the difficulty of unambiguously identifying unwinding events in NLSM
simulations of non-symmetric field configurations. The naive approach
to this problem is shown to be flawed in several ways, although the
consequences for work in which it is used \cite{PST,CPT} are unclear.
It would certainly be possible to examine the effect of adopting
inappropriate unwinding criteria on our single-texture simulations.
However because of the fundamantal differences --- particularly in the
grid resolution --- between these and the many-texture simulations in
which such unwinding criteria are used it would be difficult to
meanigfully extrapolate any effect from one to the other.

Working subject to these constraints we have been able directly to
compare the evolution of a set of initial field configurations in each
approach. In every case the fields are very strongly correlated, with
what little discrepancy there is being almost entirely radial and
simply reflecting the ability of the XCORE field to leave the vacuum
manifold. Moreover, even where such discrepancies occur the subsequent
return to near-perfect correlation indicates that the simulations are
stable to them. Given this consistency it is not surprising that,
insofar as we are able to identify them and to the accuracy of the
identification schemes, all the unwinding and non-unwinding events
occur at the same grid places and times in each approach. The fact
that the fields remain so well correlated throughout the simulations
also implies that the temporal derivatives of the field are equally
consistent. Since the evolution of the velocity field that generates
the CMB anisotopies their consistency --- both absolutely, over
individual simulations, and even more so statistically, over sets of
simulations --- is likewise a reflection of the correlation of the
velocity fields.

In conclusion we find that, subject to using a sufficiently fine
lattice and a careful handling of the discontinuity inherent in the
NLSM approach at unwinding, the XCORE and NLSM approaches to the
simulation of the evolution of global texture are entirely consistent
with one another, and hence may both be taken to give good
representations of the true field evolution.

\section*{Acknowledgements}
The author acknowledges the support of the SERC, and wishes to thank
Ed Copeland, Andrew Liddle, Albert Stebbins and Shoba Veeraraghavan
for helpful discussions.

\frenchspacing

\newpage
\section*{Figure Captions}

\vspace{30pt}
\noindent
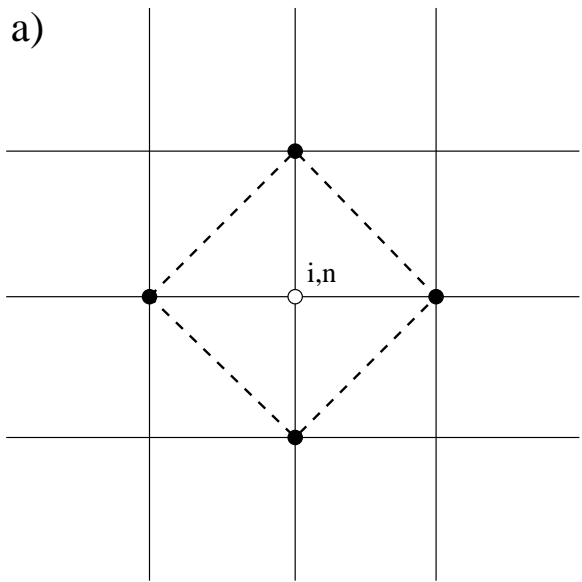
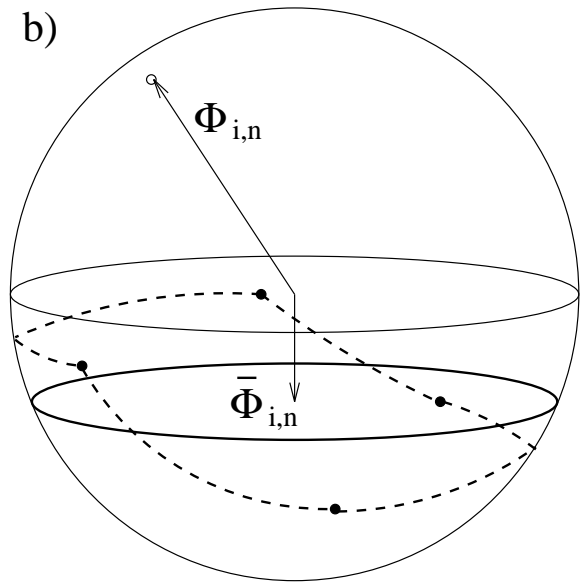
{\em Figure 1}\\
\noindent
a) Working on a 2-dimensional lattice consider each gridpoint i (empty
dot) and its 4 nearest neighbours (filled dots) at time step n.\\
\noindent
b) Mapping the gridpoint and its neighbours onto the manifold the
bisecting boundary is constructed by connecting the neighbours along
geodesics (dotted line). An unwinding should now be re-introduced
whenever the field at the given gridpoint $\Phi_{i,n}$ lies in the
larger sub-section of the thus bisected manifold. The approximation
considered here is now to replace the actual boundary with the
`average' boundary given by the intersection of the manifold with the
plane perpendicular to the mean value of the neighbours
$\bar{\Phi}_{i,n}$ (heavy line). Each boundary bisects the manifold
into two distinct sub-sections.

\vspace{30pt}
\noindent
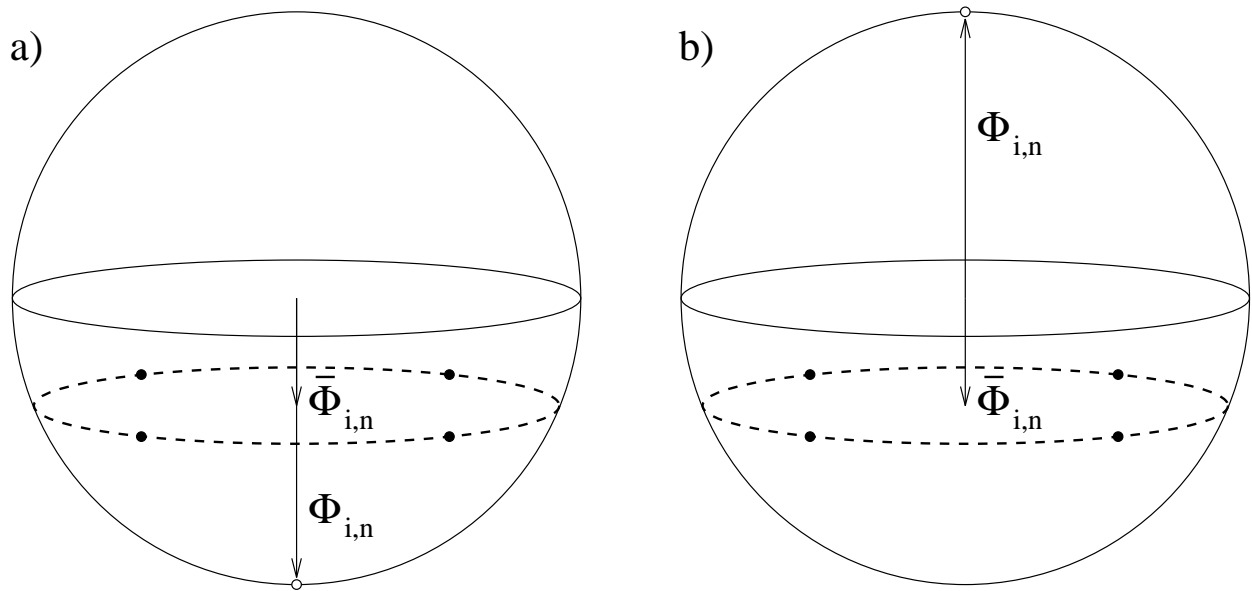
{\em Figure 2}\\
\noindent
For sufficiently (for example spherically) symmetric configurations
$\bar{\Phi}_{i,n}$ is always parallel to $\Phi_{i,n}$, so $\Phi_{i,n}
\cdot \bar{\Phi}_{i,n} < 0$ if and only if $\Phi_{i,n}$ lies in the
larger manifold sub-section --- a) no unwinding, b) unwinding. In this
case $\Phi_{i,n}$ and $\bar{\Phi}_{i,n}$ being in different
hemispheres is exactly equivalent to $\Phi_{i,n}$ lying in the larger
sub-section.

\vspace{30pt}
\noindent
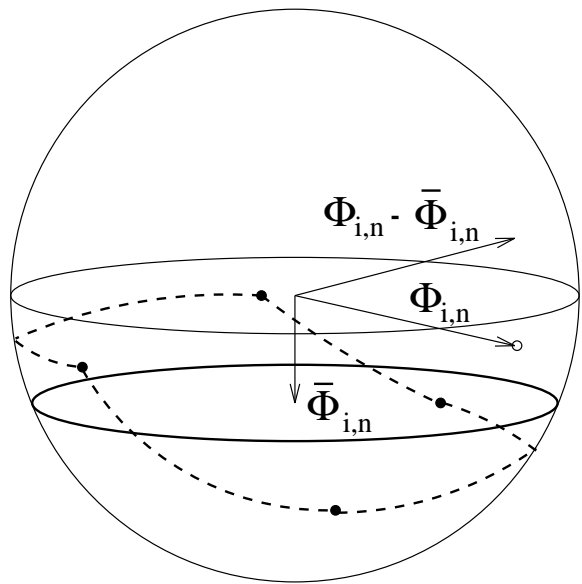
{\em Figure 3}\\
\noindent
In general $\bar{\Phi}_{i,n}$ is not parallel to $\Phi_{i,n}$, and the
equivalence noted in the symmetric case breaks down. Now $\Phi_{i,n}$
can be in the same hemisphere as $\bar{\Phi}_{i,n}$ whilst remaining
in the larger manifold sub-section. However replacing $\Phi_{i,n}$
with $\Phi_{i,n} - \bar{\Phi}_{i,n}$ restores the equivalence, so the
corrected unwinding criterion becomes $(\Phi_{i,n} - \bar{\Phi}_{i,n})
\cdot \bar{\Phi}_{i,n} < 0$.

\vspace{30pt}
\noindent
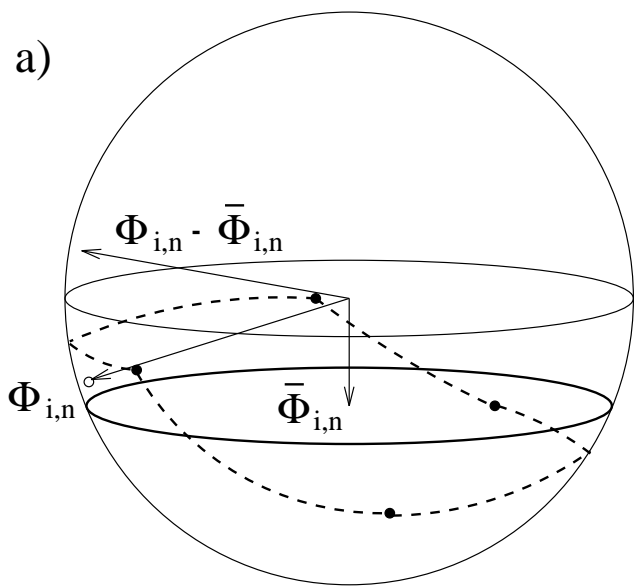
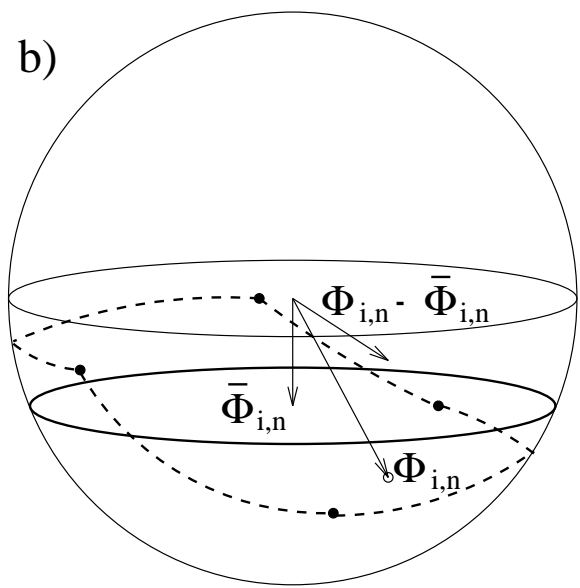
{\em Figure 4}\\
\noindent
The approximation of replacing the true boundary with the `average'
boundary breaks down whenever $\Phi_{i,n}$ lies between the two.\\
(a) if $\Phi_{i,n}$ is in the smaller true sub-section but the larger
`average' sub-section then an unwinding will be mistakenly included.\\
(b) if $\Phi_{i,n}$ is in the larger true sub-section but the smaller
`average' sub-section then an unwinding will be missed.

\vspace{30pt}
\noindent
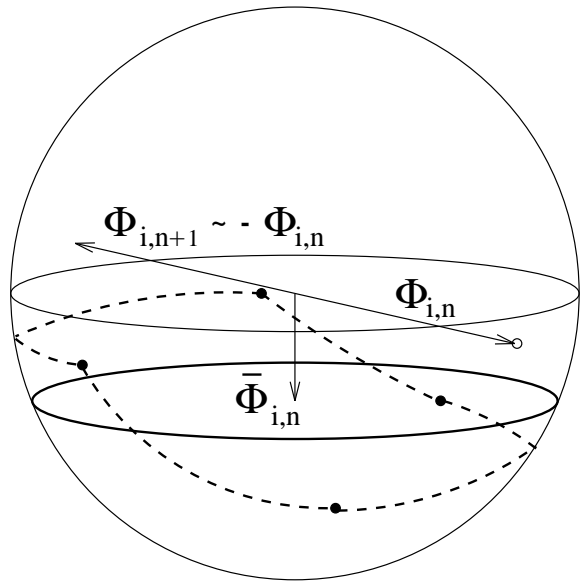
{\em Figure 5}\\
\noindent
The unwinding of a point other than the exact unwinding site in this
prescription, where the field is unwound by (to first order) being
inverted, $\Phi_{i,n+1} \sim - \Phi_{i,n}$, is clearly incorrect. For
example there are configurations in which, if the neighbouring points
do not evolve significantly, the unwound field remains in the larger
manifold sub-section, and so re-unwinds.

\vspace{30pt}
\noindent
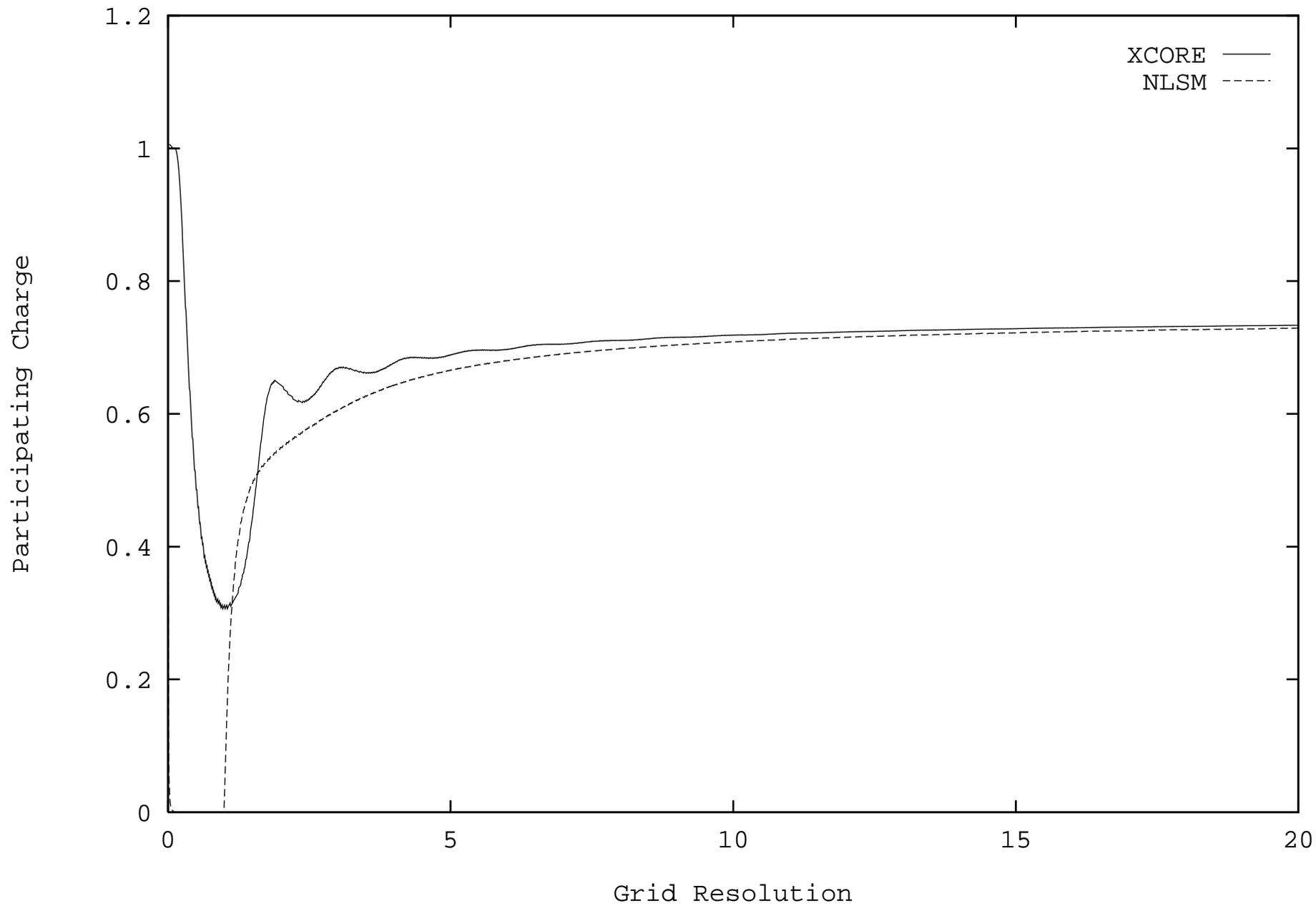
{\em Figure 6}\\
\noindent
The variation in the topological charge $Q$ causally associated with
the unwinding of a spherically symmetric texture field configuration
with the grid resolution parameter $N_{l}$ in both the XCORE (solid
line) and NLSM (solid line) approximations.

\vspace{30pt}
\noindent
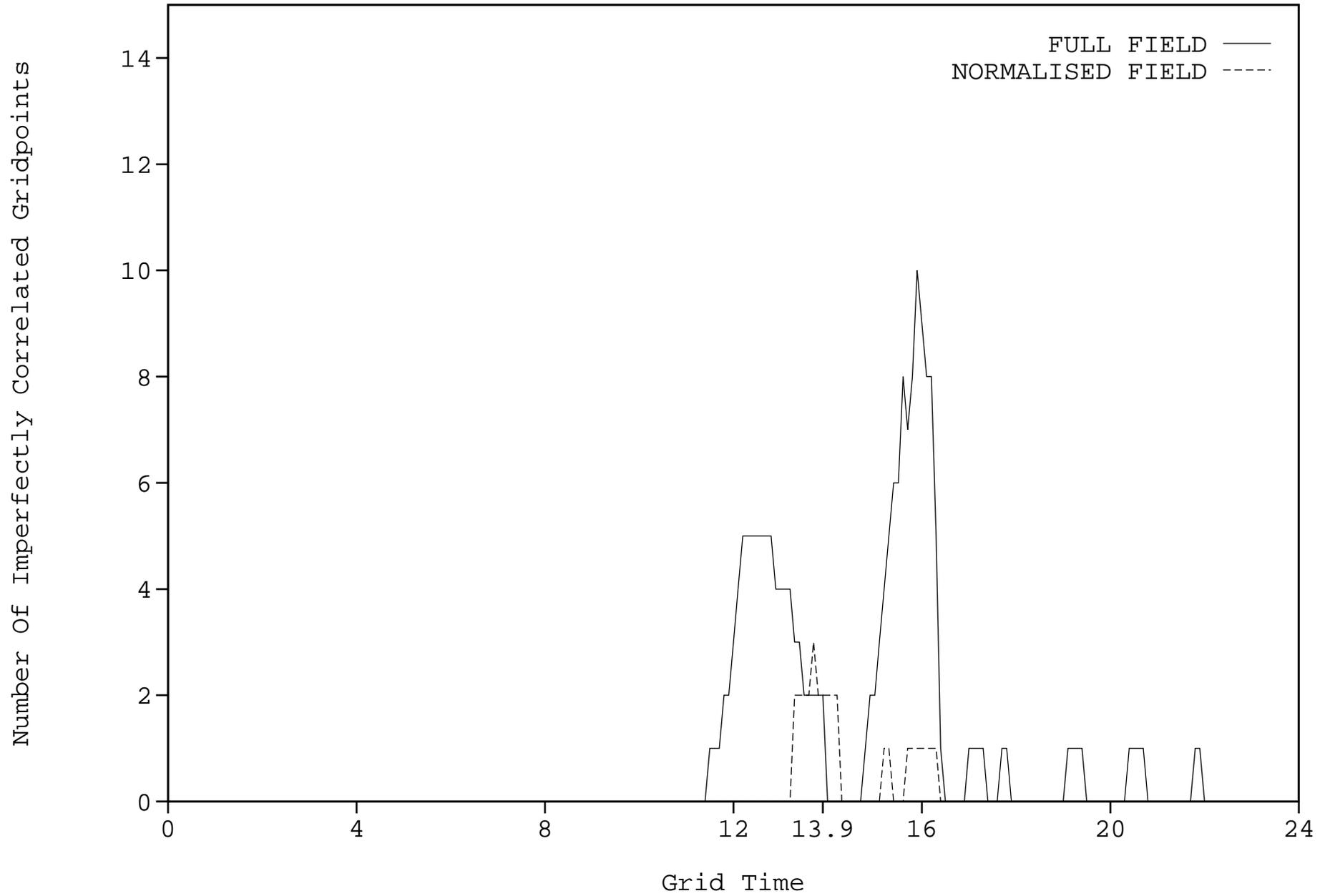
{\em Figure 7}\\
\noindent
The variation in the number of gridpoints less than 90\% correlated
between the XCORE and NLSM simulations of a typical random field
configuration with run time. Both the full fields (solid line) and the
normalised fields (dashed line) are compared. An unwinding event
occurs in at grid time 13.9.

\end{document}